\documentclass[showpacs, prb,reprint, twocolumn]{revtex4-1}

\usepackage{graphicx}
\usepackage{textcomp}
\usepackage{amsmath, wasysym}
\usepackage{longtable, multi row, threeparttable}
\usepackage{fancyhdr}
\usepackage{multirow}
\usepackage{amsfonts}
\usepackage{amssymb}

\def\gcc{$ g/cm^3 $}

\begin{document}
\headheight 50pt

\preprint{Z-GROUP: pre-print to be submitted to Physics Rev. B}

\title{Ethane-xenon mixtures under shock conditions}

\author{Rudolph~J.~Magyar}
\author{Seth~Root}
\author{Kyle~Cochrane}
\author{Thomas~R.~Mattsson}
\author{Dawn~G.~Flicker}
\affiliation{Sandia National Laboratories, Albuquerque, New Mexico, 87185, USA.}


\begin{abstract}
Mixtures of light elements with heavy elements  are 
important in inertial confinement fusion and planetary science. 
We explore the physics of molecular scale mixing through  a validation study of equation of state (EOS) properties.  
Density functional theory 
molecular dynamics (DFT-MD) at elevated-temperature and pressure is used to obtain the 
thermodynamic state properties of pure xenon, ethane, and various compressed mixture compositions along their principal Hugoniots. To 
validate these simulations, we have performed shock compression experiments using the Sandia Z-Machine. 
A bond tracking analysis correlates the sharp rise in the Hugoniot curve with the completion of dissociation in ethane.  The DFT-based 
simulation results compare well with the experimental data along the principal Hugoniots and are used to 
provide insight into the dissociation and temperature along the Hugoniots as a function of mixture composition.
\end{abstract}

\date{\today}

\maketitle


\section{Introduction}

Understanding the behavior of mixtures under intense dynamic loading conditions is needed for 
designing inertial confinement fusion (ICF) targets and developing planetary formation 
models.\cite{LHR9,WM10,OAD6}  The complicated computer simulations of these dynamic events 
rely on accurate equation of state models (EOS) that span a broad 
range of temperatures, pressures, and densities.  During impact/implosion situations or in planetary mix layers where convection occurs, pure materials 
will often dynamically mix, and a new model for the mixed material is required.  The most 
extreme limit of mixing is full homogenization on the molecular scale. We 
examine the case of thermally equilibrated and homogenized molecular mixing.  This is the final 
stage of a dynamic mix and where one would expect the most significant interspecies 
interactions resulting in the largest deviations from models based on isolated species.

Even in the case of binary mixtures, it is often impractical and expensive to perform 
experiments on many mixture compositions.  Computer simulations can more expeditiously be performed on a 
much larger set of mixture compositions at a vastly reduced cost.  However, the physics of 
mixtures even at the molecular scale is complex and the reliance on these simulations must be 
tempered with validation studies.  We report here a detailed validation study using
density functional theory - quantum molecular dynamics (DFT-QMD) simulations of warm dense matter mixtures and experiments.

As a specific example, we study the miscible cryogenic liquid mixtures of xenon and ethane (C$_2$H$_6$).  This combination is 
amenable to experiments because mixtures can be attained with modest pressures and liquid nitrogen cooling.  The mixture is also a good proxy for the polymer 
liners and heavier elements that might be used in fuel capsules on the National Ignition 
Facility and other Inertial Confinement Fusion (ICF) targets.\cite{BienerNF2012, ClarkPP2010} 
Previous studies of the isolated species have found good agreement between experiment and 
theory.  Since we expect significant interspecies interactions for molecular scale mixtures, it 
is not immediately obvious that computations of the mixtures would be as accurate as 
pure species simulations.  We perform DFT-QMD analysis of various ethane-xenon mixture ratios and calculate their Hugoniots.  The results are compared to Hugoniot 
data obtained from flyer plate experiments using Sandia's Z-Machine.  

Once validated, simulations can provide additional insight into the physical properties of 
materials. For shock compression experiments, temperature is difficult to measure.  This quantity however comes naturally in the DFT-QMD simulations, and a validation 
study of the pressures and compression ratios adds support that the predicted temperatures 
are reliable.  An important feature in the Hugoniot data of reactive systems is the onset and completion of molecular 
dissociation.  Often at the onset of dissociation, a change in slope or a plateau is observed in the Hugoniot data.  Dissociation increases the pressure and free electron density causing the Hugoniot to show significant steepening upon completion where small changes in the shock compressed density result in precipitous increases in the pressure.\cite{DickJCP1970, Mattsson2010PMP, ShermanMethane, RootCO2:2013}   However, a possible competing effect is exotic bonding 
such as the formation of xenon-hydride under pressure.\cite{Somayazulu2010}  Through a bond tracking analysis we 
can measure the stoichiometric ratios of various products to characterize the interspecies 
chemistry along the Hugoniot.


This paper is organized into 3 sections.  In the first, we discuss some of the details and 
convergence criteria used in the DFT-QMD simulations.  The second section describes the details 
of the experimental setup on the Z-Machine. and the third section summarizes the combined results of 
simulations and experiments.


\section{Density functional / quantum molecular dynamics simulations} \label{s:method}

In DFT/QMD, a number of nuclei
are moved on the Born-Oppenheimer potential-energy surface of thermally excited
electrons. The DFT method is outlined in Refs.~\citenum{HK64,KS65,MerminFT}.
The DFT/QMD simulations were performed with VASP
5.2,~\cite{vasp1,vasp2,vasp3} a plane-wave, periodic-boundary-conditions code that employs projector augmented-wave (PAW)
core functions.\cite{B1994paw,KJpaw1999}  We use stringent convergence criteria.\cite{MSD5}
The simulations are performed in the NVT ensemble (fixed number of atoms and fixed volume/density at
prescribed temperature).  We employ velocity scaling as the thermostat for the simulations; 
however, additional simulations using an Nose-Hoover thermostat show negligible difference for Hugoniot states.
Complex k-point sampling with the Balderashi mean-value point is applied because its accuracy and 
efficiency for disordered structures at high temperature. We run Mermin's
finite temperature formulation of DFT with ground-state exchange
correlation functionals,\cite{MerminFT} shown to be critical
for high energy-density applications.\cite{MPDdeuterium2003}  We report the
results for only AM05,~\cite{AM05,MAP2008}
which is particularly well suited to describe compressed solids and liquids.\cite{RMC10} Results
within the local density approximation (LDA) are comparable.

A direct route to compare experimental shock data to DFT-MD simulations is
through the calculation of Hugoniot states.  The hydrostatic Hugoniot
condition can be expressed as $2(E-E_{ref})=(P+P_{ref})(V_{ref}-V)$ where E
is the internal energy per mass, P is the system pressure, and V is the specific volume and is related to the inverse of the mass density, $V= 1 / \rho$.  The subscript \textit{ref} refers to the reference state, which is at the initial conditions of the experiment.

Each simulation was allowed to equilibrate at a
constant temperature and density for multiple picoseconds or until the block
averaged~\cite{allen:1987} standard deviation of the mean was less than
$1\%$. At each density, we used two temperatures to approximate the Hugoniot
relation; one temperature such that the pressure and energy were too high
and the other too low.  We then interpolated between them to obtain the
Hugoniot pressure, energy, and temperature.  For higher compression points along the Hugoniot, an alternative approach is possible in which several densities at fixed temperatures are performed and used to locate the Hugoniot state.

Throughout, we reserve $x$ for the mass mixing ratio defined as 
\begin{equation}
x = \frac{n_{ethane}M_{ethane}}{n_{Xe}M_{Xe}+n_{ethane}M_{ethane}}
\end{equation}
where n is the number density of molecules (atoms) and M is the molecular mass of each species.  This form is convenient because EOS tables are often given in terms of mass densities.  



\begin{figure}[ht]
\centering
\includegraphics[width=3.0in]{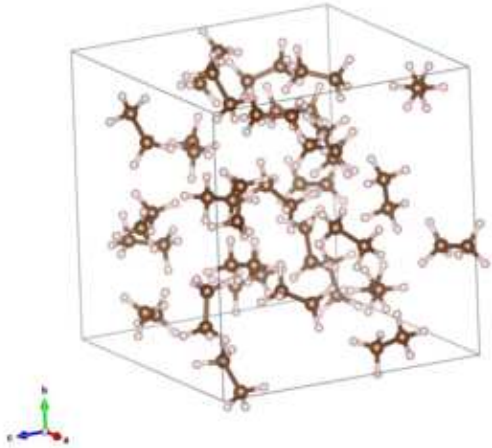}
\caption{
Snapshot of the liquid ethane reference state nuclear positions within a super-cell at density $\rho=0.571$ \gcc at 163 K from a DFT/QMD  AM05 calculation.
Carbon atoms are brown and hydrogen atoms are white. }
\label{f:ethref}
\end{figure}

\begin{figure}[ht]
\centering
\includegraphics[width=3.0in]{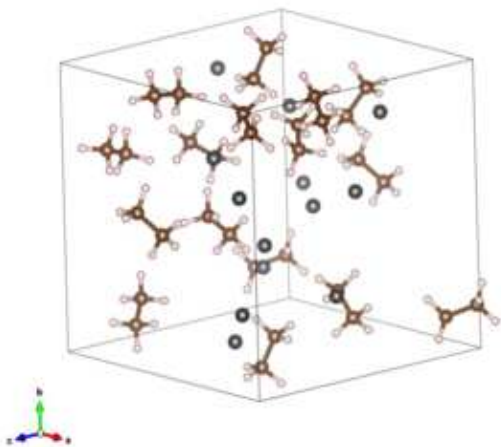}
\caption{Same as Fig. \ref{f:ethref} except now including Xenon with mass ratio x=0.19 and density $\rho=1.676$ \gcc. Xenon atoms are grey. }
\label{f:xeethref}
\end{figure}

The reference state ($E_{ref}$, $P_{ref}$, and $V_{ref}$) for ethane, Fig. \ref{f:ethref}, was
chosen to closely match experimental initial conditions of 0.571 \gcc and 163 Kelvin determined from cryomixture data.\cite{Filipe2000}  For the
50/50 molar mixing $x=0.19$, the reference state at 163K and P $< 10$ kBar is $\rho_{ref}=1.676$ \gcc and is shown in Fig. \ref{f:xeethref}.
For the 50/50 mass mixing $x=0.5$, the reference state at 163K and P $< 20$ kBar is $\rho_{ref}=0.960$ \gcc.  The densities are taken to be given for the reference state and we do not perform calculations to optimize over density at ambient pressure. The results for pure xenon are reported in a previous validation study.\cite{RMC10}

A computer code was created to position atoms within a super cells to
represent the Xe-Ethane mixture.  The center of mass positions of the ethane
molecules were chosen to uniformly fill a super cell with $2^3$,
$3^3$, and
$4^3$ ethane molecules.  To achieve a desired mix ratio $x$, a
number of ethane molecules were randomly substituted with Xe.  It was found that the
largest super cell ($4^3$)was impractical for more than a few simulations. The $3\times3\times3$ was
converged with respect to simulation size through a scaling test of the pressure at 1 MBar.
The cell size was scaled to achieve a given density.  Note that changing the
density was done adiabatically by gradually scaling the super cell, running
a dynamic DFT-MD simulation to equilibrate the bond lengths, and then
scaling further.  A sudden large density scaling would strain bonds, increasing the system energy
too dramatically and result in premature bond breaking.

We determined that 27 ethane molecules in each super-cell is sufficient by simulating 64 molecules at 0.8, 1.1, 1.2, 1.5, and 1.8 \gcc and comparing to 27 molecule results for pressure and energy.  The large super-cell simulations gave the same Hugoniot points to within $\leq 0.5 \%$ in pressure.
The number and type of atoms in each super-cell varies according to the mix ratio.  The 50/50 mass mix ($x=0.5$) had 5 xenon and 22 ethane molecules (5 Xe, 44 C 132 H) for a total of 181 atoms.  The 50/50 molar mix ($x=0.19$, see Fig.~\ref{f:xeethref} had 13 Xenon and 13 Ethane (13 Xe, 26 C and 78 H ) for a total of 117 atoms.

We address the degree of dissociation using a bond tracking analysis on the nuclear positions.\cite{RootCO2:2013}  Two elements are considered bonded if they are within a certain bonding radius for at least as long as a given persistence time.  For the results quoted here, we choose a persistence time of 100 fs which corresponds to an inverse vibrational frequency time scale for the C-C bond.  The bond lengths we target are slightly longer than the equilibrium bond lengths as we expect elongation with temperature.  
We use Xe-Xe 0.25 \AA, Xe-C 2.0  \AA, Xe-H 2.0  \AA, C-C 1.68  \AA,  C-H 1.28  \AA, and  H-H  0.8 \AA.  Since Xe-Xe bonding is unlikely, we chose an arbitrarily small bond cut-off.


Results from the DFT simulations are listed in Tables~\ref{t:ethanehug}-\ref{t:xeethmolar}  and are compared to the experimental data in Section~\ref{SimsResults}.

\begin{table}
\caption{DFT-MD results for the liquid ethane Hugoniot}
\label{t:ethanehug}
\centering
\begin{tabular}{c c c c}
\hline\hline
Density & Pressure & Energy & Temperature \\
($g/cm^3$) &  (GPa)  & (eV/atom) & (K) \\
\hline
0.571   	& 0.4674   &     -5.264	& 163 \\
0.800   	& 2.290 	& 	-5.238 	& 204 \\
1.00  	& 7.438 	& 	-5.137  	& 530 \\
1.10   	& 11.23 	&	-5.054  	& 618 \\
1.20 		& 16.92   	& 	-4.918	& 947 \\
1.30		& 25.30	&	-4.705	& 1487 \\
1.40		& 35.44	&	-4.471	& 2175 \\
1.50		& 46.98	&	-4.232	& 2876 \\
1.60		& 55.71	& 	-3.953	& 3142 \\
1.70		& 64.50	&	-3.677	& 3386 \\
1.80		& 78.20	& 	-3.259	& 4043 \\
1.906	& 110.3	& 	-2.586	& 7000 \\ 
1.929	& 134.7 	&  	-2.019  	& 10000 \\
1.942	& 175.0	&	-1.040	& 15000 \\
1.95 		& 187.4 	&	-0.754	& 16669 \\
1.964	& 215.8	&	-0.056	& 20000 \\ 
1.993	& 300.8	&	2.044	& 30000 \\ 
2.000	& 320.1	&	2.526	& 32139 \\ 	
2.019	& 390.9	&	4.288	& 40000 \\ 

\hline\hline
\end{tabular}
\end{table}

\begin{table}
\caption{DFT-MD Hugoniot points for the 50/50 mass mixture ($x=0.5$) of liquid xenon-ethane}
\label{t:xeethmass}
\centering
\begin{tabular}{c c c c}
\hline\hline
Density & Pressure & Energy & Temperature \\
($g/cm^3$) &  (GPa)  & (eV/atom) & (K) \\
\hline
0.960  &  0.893     &  -5.151  &    163 \\
1.250  &  2.068   &   -5.098  &    214  \\
1.500  &  4.270    &  -5.076  &    226 \\
1.600  &  5.875    &  -5.043  &    308 \\
1.750  &  9.055    &  -4.971  &    495 \\
2.000  &  16.20   &  -4.790  &   1003 \\
2.250  &  28.25   &  -4.465  &   1972 \\
2.500  &  42.81   &  -4.041  &   2919 \\
2.750  &  58.30   &  -3.560  &   3730 \\
2.997  &  77.51   &  -2.948  &   5000 \\
3.000  &  77.78   &  -2.939  &   5021 \\
3.050  &  83.12   &  -2.770  &   5451 \\
3.140  &  117.0 &  -1.753  &  10000 \\
3.191  &  186.2  &  0.289   &  20000 \\
3.231  &  260.2  &  2.492   &  30000 \\
3.307  &  337.9  &  4.866   &  40000 \\
3.360  &  421.4  &  7.418   &  50000 \\
3.378  &  508.0  &  10.03  &  60000 \\
\hline
\hline\hline
\end{tabular}
\end{table}

\begin{table}
\caption{DFT-MD Hugoniot points for the 50/50 molar mixture ($x=0.19$) of liquid xenon-ethane}
\label{t:xeethmolar}
\centering
\begin{tabular}{c c c c}
\hline\hline
Density & Pressure & Energy & Temperature \\
($g/cm^3$) &  (GPa)  & (eV/atom) & (K) \\
\hline
1.676  &  0.780     &  -4.708  &   163 \\
2.500  &  3.257    &  -4.664  &   167 \\
2.600  &  3.997    &  -4.646  &   209 \\
2.800  &  5.983    &  -4.594  &   310 \\
3.050  &  8.876    &  -4.508  &   500 \\
3.250  &  10.64   &  -4.445  &   800 \\
3.500  &  16.23   &  -4.264  &  1221 \\
3.750  &  21.92   &  -4.062  &  1834 \\
4.000  &  29.58   &  -3.783  &  2579 \\
4.500  &  44.98   &  -3.173  &  3652 \\
5.000  &  67.10   &  -2.266  &  5902 \\
5.250  &  81.74   &  -1.655  &  7409 \\
5.514  &  128.0  &  0.198  &  15000 \\
5.550   &  142.8   &  0.786  &   17463  \\
5.609  &  156.1  &  1.327  &  20000 \\
5.742  &  219.0  &  3.855  &  30000 \\
5.850  &  286.2  &  6.576  &  40000 \\ 
6.089  &  442.1   & 13.10  & 60000 \\ 
\hline
\hline\hline
\end{tabular}
\end{table}

\section{Experimental Approach and Results}

\begin{figure}
\includegraphics[width=2.0in]{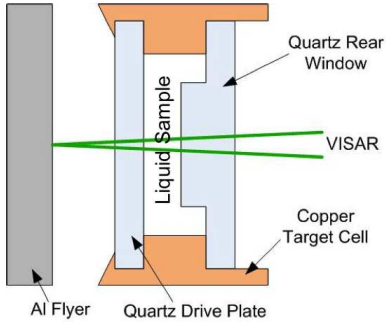}
 \caption{Schematic view of the target cell configuration for the ethane and the ethane-xenon mixture shock compression experiments}
 \label{ExptSchematic}
\end{figure}

To compare with the DFT simulations and to examine pure ethane and ethane-xenon mixtures at extreme conditions,
we performed a series of shock and reshock experiments using Sandia's Z-Machine.\cite{ZMachRef}  The Z-Machine is capable of accelerating solid aluminum flyers up to velocities of 40~km/s.\cite{LemkeZFlyer05}   This technique has been used to accurately measure the Hugoniot and second shock states for several cryogenic fluids such as deuterium, \cite{MDKD2:2001, MDKD2:2004}  xenon,\cite{RMC10} and CO$_2$\cite{RootCO2:2013}.  The target cell consists of two z-cut $\alpha$ quartz windows on the front and back of the target cell.   A copper spacer is placed between the quartz windows.  The rear window consists of two quartz windows bonded together using Epotek 301-2 to form a top-hat.  The combination of the copper spacer and the smaller diameter quartz window in the top-hat set the cryogenic liquid sample thickness.  The quartz windows are anti-reflected coated to reduce Fresnel reflections at the interfaces. The target cells are connected to a liquid nitrogen cryostat\cite{RMC10, RootCO2:2013} and cooled to temperature.  A schematic view of the cryotarget and experimental configuration is shown in Figure~\ref{ExptSchematic}.  Further details of the cryotarget and system can be found in Ref.~\citenum{RMC10}.

Velocity interferometry (VISAR) \cite{BarkerVISAR} is the primary diagnostic for the shock experiments.  The Sandia VISAR system consists of two dual velocity per fringe (vpf) interferometers, which allows up to four different vpf settings to be used on a single target.  This eliminates 2$\pi$ ambiguity in the data analysis.  Typical uncertainty in the velocity records is $<0.5$\%. As shown in the experimental schematic view (Fig.~\ref{ExptSchematic}), the 532~nm laser used for the VISAR passes through the target cell and reflects off the aluminum flyer.  The velocity of the aluminum flyer is measured up to impact with the quartz front window.  The shock front generated in the quartz window at impact is reflective\cite{MDKQuartz09} and the shock velocity in the quartz is measured directly using the VISAR.  The shock transmitted into the ethane or the ethane-xenon mixture has a reflective shock front allowing the shock velocity in the cryogenic liquid sample to be measured directly.  Lastly, the shock front in the rear quartz top-hat is also reflective, from which the state in the quartz and the reshock state in the liquid sample can be determined accurately.

The target cells were filed with high purity ($>$99.99\%) ethane gas (Matheson TriGas) to 16.8 PSI and cooled to the final temperature.  The initial density of liquid ethane was determined from the fit to the experimental data in Reference~\citenum{Haynes1977} with an uncertainty of 0.5\%.  The index of refraction of liquid ethane was determined from the data in Reference~\citenum{Weber1976}.  Ethane-xenon gas mixtures were supplied by Matheson TriGas and the molar ratio content was verified using mass spectrometry.  The initial density of the ethane-xenon mixtures is determined from a linear fit to the experimental density data in Reference~\citenum{Filipe2000}.  The index of refraction for the mixture was calculated using the Lorentz - Lorentz mixing rule: 
\begin{equation}
  \frac{n_{12}^2-1}{n_{12}^2+2} = \Phi_{1}\frac{n_{1}^2-1}{n_{1}^2+2} + \Phi_{2}\frac{n_{2}^2-1}{n_{2}^2+2}
  \label{LLRule}
\end{equation}
with
\begin{equation*}
\Phi_{i} = y_{i}m_{i}/\rho_{i}  / \sum_j{y_{j}m_{j}/\rho_{j}}.
\end{equation*}
where $y$ is the molar fraction, $m$ is the atomic mass, and $\rho$ is the density.\cite{Mehra2003}  The initial density and index of refraction values for liquid xenon were taken from the literature.\cite{Leadbetter1965}  Since the shock velocity is measured directly, the measured shock velocities can be integrated with respect to time to ensure the distance traveled is consistent with the sample size.

The principal Hugoniot state is derived using a Monte Carlo impedance matching with a cubic fit to the published quartz data\cite{MDKQuartz09, RootCO2:2013} and a quartz release model based on an effective Gr\"unesein $\Gamma$ parameter.\cite{MDKQuartzRel2013}  For the experiments on the mixtures, the uncertainty in the molar concentration contributes primarily to the uncertainty in  initial density (up to $\approx$4\%), while the effect on the index of refraction is small, $<0.05$\%.  The small effect on the index is primarily caused by xenon and ethane having similar indices of refraction at the experimental temperatures: at 161.5~K $n_{Xe}$ = 1.392 and $n_{ethane}$ = 1.343.  Tables~\ref{tab:EthaneData} - \ref{tab:MolarMixData} list the experimental observables ($U_S^{Quartz}$ and $U_S^{Liquid}$) and the derived Hugoniot states. 

\begin{table*}[ht]
  \addtolength{\tabcolsep}{6pt}
\renewcommand{\arraystretch}{0.7}
\caption{Experimental data of the principal Hugoniot for shock compressed liquid ethane.}
\begin{tabular}{cccccccc}
\hline
   Shot & Quartz $U_S$   &    $\rho_0^{Ethane}$   & $T_0$ &   $U_P$  &  $U_S$   &  $\rho$   		&  P  \\ 
            &         (km/s)       &       (g/cm$^3$)      &    (K)   &    (km/s)  &   (km/s)   &  (g/cm$^3$)    & (GPa)  \\ \hline\hline
Z2331 & 17.32$\pm$0.03 &  0.571 		      &  162.0    & 14.00 $\pm$ 0.05    &   20.07  $\pm$ 0.03 & 1.888  $\pm$ 0.017   &     160.4 $\pm$  0.8 \\
Z2226 & 18.17$\pm$0.04&   0.570  		      &  161.5 & 14.89 $\pm$ 0.06    &   21.3   $\pm$ 0.07  & 1.897  $\pm$ 0.024   &     181.1 $\pm$  1.1  \\
Z2277 & 20.52$\pm$0.04&   0.572		      &  163.5 & 17.40 $\pm$ 0.06    &   24.6   $\pm$ 0.06  & 1.952  $\pm$ 0.021   &     244.5 $\pm$  1.4 \\
\hline
\end{tabular}
\label{tab:EthaneData}
\end{table*}

\begin{table*}[ht]
  \addtolength{\tabcolsep}{6pt}
\renewcommand{\arraystretch}{0.7}
\caption{Principal Hugoniot experimental data for shock compressed liquid Xe-ethane $\sim$50/50 mass mix ($x=0.5$).}
\begin{tabular}{cccccccc}
\hline
   Shot    & Quartz $U_S$		&  $\rho_0^{MIX}$		& $T_0  $    &  $U_P$    		&  $U_S$      &  $\rho$            &  P  \\ 
               &    (km/s)         		&     (g/cm$^3$)   		&    (K)        &     (km/s) 		&     (km/s)     &  (g/cm$^3$)   &   (GPa)   \\  \hline\hline
Z2527N  & 23.25 $\pm$ 0.07	& 	0.958$\pm$0.021	&   161.5     &18.61 $\pm$ 0.13 & 25.75  $\pm$ 0.07 & 3.457 $\pm$ 0.061 & 459.1 $\pm$ 8.1 \\
Z2527S  & 24.23 $\pm$ 0.08  	& 	0.958$\pm$0.021	&   161.5     &19.59 $\pm$ 0.14 & 27.08 $\pm$ 0.09  & 3.464 $\pm$ 0.067  & 508.1 $\pm$ 9.0 \\
\hline
\end{tabular}
\label{tab:MassMixData}
\end{table*}

\begin{table*}[ht]
  \addtolength{\tabcolsep}{6pt}
\renewcommand{\arraystretch}{0.7}
\caption{Experimental data of the principal Hugoniot for shock compressed mixtures of a 50/50 molar mixture ($x=0.19$) of liquid ethane and liquid xenon with an initial density of $\rho$=1.676 $g/cm^3$ . }
\begin{tabular}{cccccccc}
\hline
  \multirow{2}{0.0in}{}
    Shot	& Quartz $U_S$ 	&  $\rho_0^{MIX}$	&  $T_0$	&       $U_P$  	       &  $U_S$               &  $\rho$ 	      &  P		 \\
    		&(km/s)      		&   (g/cm$^3$)  	&      (K)	&	(km/s) 	       &   (km/s)             &       (g/cm$^3$)      &  (GPa)   \\  \hline\hline
Z2226   	& 18.15 $\pm$ 0.05	&  1.676$\pm$0.073	&  163.5	&  12.18  $\pm$  0.16  &  17.12  $\pm$  0.09  &  5.814  $\pm$ 0.142 &  349.4 $\pm$  11.3 \\
Z2277    	& 20.56 $\pm$ 0.04	&  1.676$\pm$0.073	&  163.5	&  14.29  $\pm$  0.18  &  19.62  $\pm$  0.08  &  6.171 $\pm$ 0.122 &  469.6 $\pm$  15.0 \\
Z2295    	& 20.95 $\pm$ 0.03	&  1.676$\pm$0.073 &  163.0	&  14.63  $\pm$  0.18  &  20.08  $\pm$  0.07  &  6.173 $\pm$ 0.109 &  491.9 $\pm$  15.7 \\

\hline
\end{tabular}
\label{tab:MolarMixData}
\end{table*}



\section{Discussion and Results}
 \label{SimsResults}

\begin{figure}
  	\includegraphics[width=3.1in]{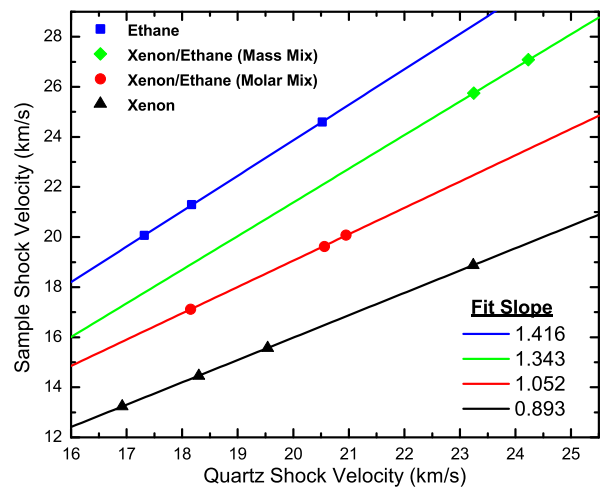}
	\caption{Experimental measurements: quartz shock velocity ($U_S^{Quartz}$) vs. sample shock velocity ($U_S$).  The data for the pure xenon is from Ref.~\citenum{RMC10}.  The lines are linear fits to the data.}
	\label{QshockVshock}
\end{figure}

Figure~\ref{QshockVshock} plots the experimental observables $U_S^{Quartz}$ and $U_S^{Liquid}$ for the four samples: pure xenon, 50/50 molar mixture ($x=0.19$), 50/50 mass mixture ($x=0.5$), and pure ethane.   The results show that the xenon $U_S^{Xenon}$ has the lowest velocity as a function of $U_S^{Quartz}$.  The observed  $U_S^{Liquid}$ increases as the amount of ethane increases.  The molar and mass mixture observable data are bounded by the pure ethane and xenon data as expected.  A large change in the slope parameter occurs in the linear fits to the data between the molar and mass mixtures.    That the change in the slope parameter occurs between the mass and molar mix ratios suggests that the high pressure response is governed more by the differing masses of the elements rather than the relative chemical interactions. 
 
\begin{figure}[ht]
\centering
\includegraphics[width=3in]{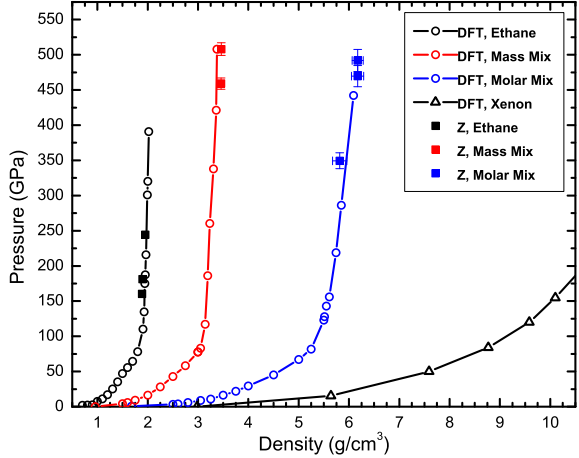}
  \caption{Hugoniot data from the Z-experiments and the DFT simulations for the pure ethane, mass mixture, molar mixture, and pure xenon.  All DFT data start from a reference state density for the system at T=163K and 0 bar pressure.  The initial states for the experimental data are listed in Tables~\ref{tab:EthaneData} - \ref{tab:MolarMixData}  Error bars for the experimental data are on the order of the symbol unless indicated otherwise.}
\label{HugDataCompiled-rhoP}
\end{figure}

Figure ~\ref{HugDataCompiled-rhoP} shows the compiled Hugoniot data sets for the pure and mixed liquids including the pure xenon, which was shown in Ref.~\citenum{RMC10}.  Tables~\ref{t:ethanehug}-\ref{t:xeethmolar} show the calculated values for points along the Hugoniot.  The temperatures given in clean values such as 40000K were fixed when a density scaling procedure is performed.  These are often the higher temperature points where the sharp rise in the Hugoniot makes temperature scaling inaccurate.  Note that for ethane, we have both density and temperature scaling points interwoven suggesting that both methods are equally accurate in an intermediate region.  The results from the DFT simulations show good agreement with the experimental data.
For the pure ethane and the two mixtures, the Hugoniot is observed to undergo a sharp steepening after the complete dissociation of the ethane.  The DFT simulation results tend to show a slightly stiffer Hugoniot than the experimental results at the highest pressures.  This has been observed in other comparisons with experiments,\cite{RMC10, RootCO2:2013} which is likely caused by the computational limits on the number of plane waves that can be used to describe the energy states.

\begin{figure}
   \includegraphics[width=3.0in]{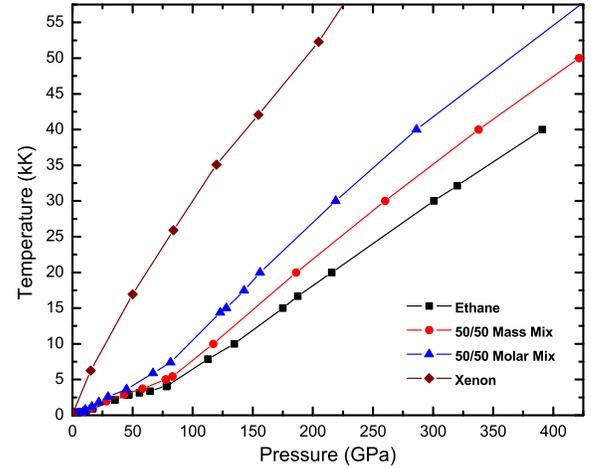}
    \caption{DFT Hugoniot calculations in pressure - temperature space.}
    \label{DFT-PTHug}
 \end{figure}

Experimentally, measuring temperature is difficult.  However, temperature is a known quantity in the DFT simulations.  Figure \ref{DFT-PTHug} shows the DFT Hugoniot data in pressure - temperature space.   For a given pressure, the shock temperature in pure xenon is noticeably higher than the pure ethane or ethane-xenon mixtures.  Dissociation of the ethane absorbs energy generated in the shock, thus keeping the temperature lower.  The addition of xenon means fewer dissociating ethane molecules to absorb the shock energy and higher temperature for a given pressure.   Because of xenon's higher mass it moves slower on average at a given temperature than ethane.  This results in less frequent collisions and resulting chemistry for a given temperature.

The temperature data (Fig.~\ref{DFT-PTHug}) also provides indication of complete dissociation of the ethane molecules around 100~GPa where the slope in the P-T data increases.  The mass mixture is mostly ethane and still exhibits the discreet regions of differing slopes in P-T before the sharp up turn. This is not evident in the molar results suggesting that the chemistry is highly affected by the 1-to-1 ratio of xenon to ethane. The inclusion of Xe also has a mollifying effect on the sharp upturn observed in ethane $\rho$-P Hugoniot (Fig.~\ref{HugDataCompiled-rhoP} ), which is a signature of dissociation completion. Both the mass and molar mixture exhibit a sharp upturn but at slightly higher pressures and the post-dissociation Hugoniot is softer than the pure ethane.    

\begin{figure}
\includegraphics[width=3in]{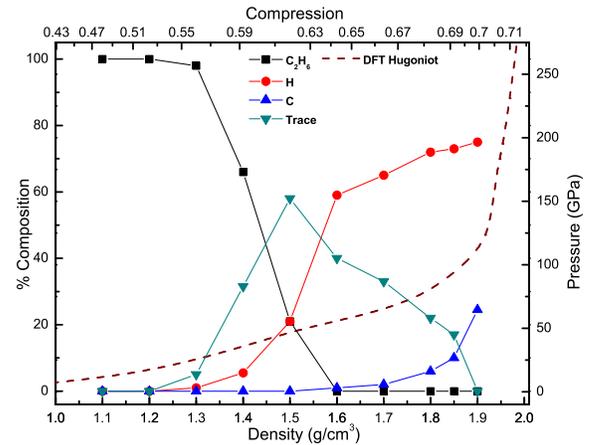}
\caption{The percent of total atoms in the system that are in each molecule type.  The xenon atom percentage is not shown.  Trace refers to all possible C-C and C-H combinations excluding C$_2$H$_6$.  Juxtaposed on the plot is  the DFT Hugoniot (dashed-line).  Complete dissociation is observed by 1.9 g/cm$^3$.}
\label{DecompEthane}
\end{figure}

In Fig. \ref{DecompEthane}, we juxtapose the principal Hugoniot for shocked liquid ethane with the decomposition pathway of the ethane determined from the DFT simulations. Ethane decomposition starts at approximately 29~GPa and corresponding density of 1.3~\gcc.  In terms of compression, $\mu = 1-\rho_{0}/\rho$; the onset of decomposition is $\mu=0.56$.  The decomposition begins with the removal of H from the ethane molecule.  Initially, the trace molecules observed are primarily C$_2$H$_5$ and C$_2$H$_4$, but can include other C-H molecules.  As the shock pressure increases, the number of liberated H and the number of trace molecules increases.  By 1.7 \gcc, most molecules have decomposed with only a few (1\% to 2\%) CH and $CH_2$ present at any given time. At pressures above 55 GPa and $\rho = $ 1.7~\gcc, the trace C$_X$H$_Y$ molecules begin to dissociate rapidly into C and H atoms.  Complete dissociation - no C$_2$H$_6$ and no trace C-C or C-H molecules occurs by 110~GPa and a density of 1.9~\gcc ($\mu = 0.70$).

\begin{figure}
  \includegraphics[width=3in]{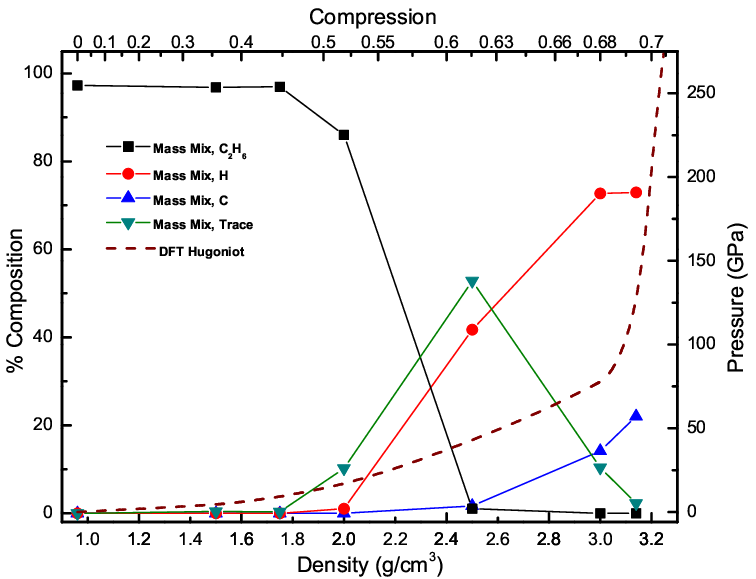}
  \caption{Decomposition of the 50/50 mass mixture ($\rho_0$=0.96g/cm$^3$, $x=0.50$) along the Hugoniot.  The dashed line indicates the DFT calculated Hugoniot.}
  \label{DecompMassMix}
 \end{figure}

Figure~\ref{DecompMassMix} plots the decomposition pathways for the mass mix ($\rho_0 = 0.960$\gcc, $x=0.50$).  The onset of appreciable decomposition starts after 9~GPa and a density of 1.75~\gcc, which corresponds to a compression of $\mu = 0.45$.  The remainder of trace molecules grows rapidly with shock pressure obtaining a maximum at 43~GPa.  Above this pressure, atomic C begins to form.  Decomposition is completed by 117~GPa and $\rho = 3.14$\gcc.  The compression at this state is $\mu = 0.69$.  At this point, the Hugoniot steepens significantly as shown in Figure \ref{DecompMassMix}.  

\begin{figure}
\includegraphics[width=3.0in]{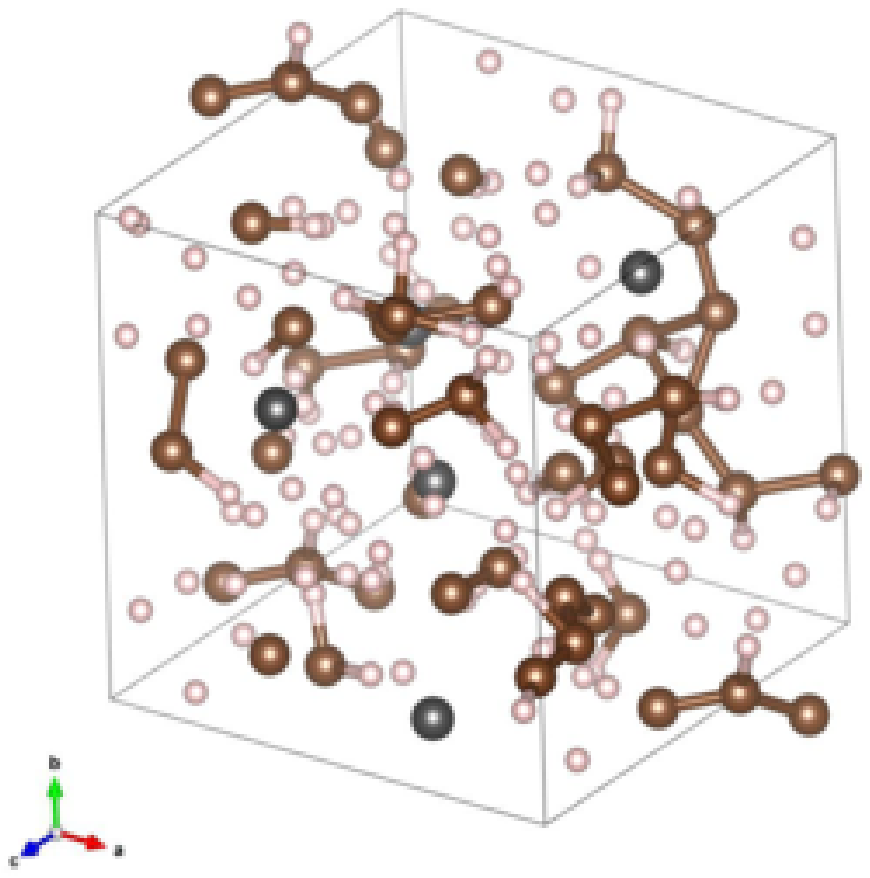}
\caption{Same as Fig. \ref{f:ethref} except now including Xenon with mass ratio x=0.19 and density $\rho=5.30$ \gcc at $T=10,000K$. Xenon atoms are grey. }
\label{f:xeethhot}
\end{figure}

\begin{figure}
  \includegraphics[width=3in]{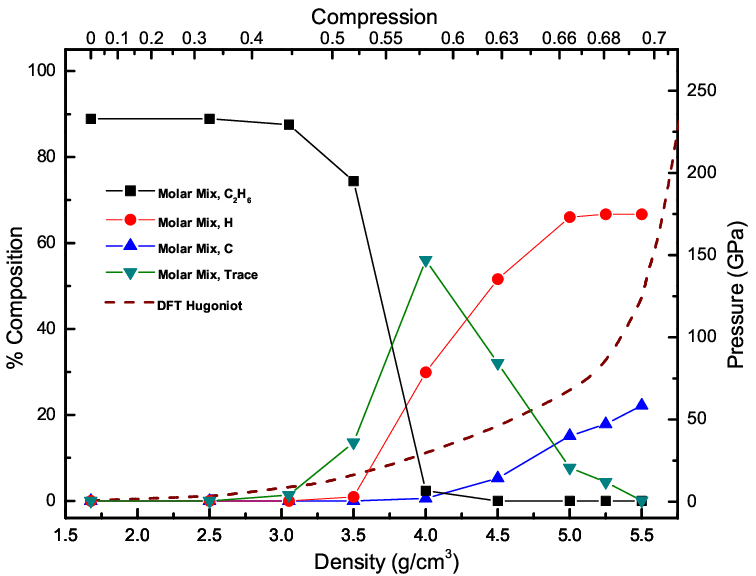}
  \caption{Decomposition of the 50/50 molar mixture ($\rho_0$=1.676\gcc, $x=0.19$) along the Hugoniot.  The dashed line indicates the DFT calculated Hugoniot.}
  \label{DecompMolarMix}
 \end{figure}

Figure~\ref{f:xeethhot} shows a super-cell of the molar mixture ($x=0.19$) near complete dissociation.  The carbon 
atoms appear to form cross-linking chains under compression with the majority of carbon atoms participating in chain formation. This is because the bonds shown are for instantaneously close carbons.  The persistence time of these chains is shorter than the 100fs window.  However, the separate species co-location suggests the tendency for H and C to demix 
under these conditions.  Figure~\ref{DecompMolarMix} shows the decomposition pathway for the molar mixture ($\rho_0 = 
1.676$\gcc, $x=0.19$). The additional xenon increases the pressure for complete dissociation to $\approx$126~GPa and 
a density of 5.5\gcc; this corresponds to a compression of $\mu$=0.69 for complete dissociation.  While the addition 
of xenon causes the complete dissociation pressure and temperature to increase slightly with increased xenon, the end 
compression state $\mu$ is the nearly same for ethane and mixtures of xenon-ethane.  In neither the mass ($x=0.5$) 
nor the molar ($x=0.19$) mixtures did we observe a xenon-hydride as observed in static high pressure experiments 
\cite{Somayazulu2010}.


The similar compression ratios indicates that complete dissociation for the pure ethane and ethane-xenon mixture systems has a strong dependence on the density.  Interestingly, the ratio $\mu=0.69$ for ethane is also similar to the compression ratios for other C-C bonded systems.  Polystyrene completely dissociates at $\mu$=0.62;\cite{WangPP2011} polyethylene is completely dissociated at $\mu$=0.62;\cite{Cochrane2011} and poly 4-methyl-1-pentene (PMP) is completely dissociated at $\mu$=0.68.\cite{Mattsson2010PMP}  This suggests that a limiting compression exists for C-C bonded systems.

\section{Conclusion}

We have performed series of experimental measurements and DFT simulations on pure ethane and xenon - ethane mixtures to better understand the response of atomic mixtures at high 
pressures and temperatures.  The experimental measurements provided accurate Hugoniot data for ethane and xenon mixtures that were used to validate DFT simulations.  Examination of the 
DFT simulations showed that ethane dissociation begins with the removal of H.  As the pressure state increases, the ethane eventually completely dissociates into C and H atoms.  The addition of xenon increased the pressure and temperature for complete dissociation, but that the ultimate compression for dissociation of the system remained the same regardless of the xenon concentration.

These results suggest that the primary role of introducing xenon into the mixture is to raise the temperature needed for dissociation and the onset of the pressure 
up-turn of the $\rho-P$ Hugoniot relative to the pure ethane result.  This result is dominated by kinematic considerations and inter-species chemistry plays a minor role in the data.  Since the upturn occurs at the similar compression ratios for each species, the state of complete dissociation is likely caused by a limiting compression rather than purely kinetic reasons.

Hydrodynamics simulations of high energy density physics phenomena require a detailed knowledge of the equation of state of the constituents as well as their mixtures. In this paper, we analyzed validate high-fidelity DFT/QMD simulations of mixtures under shock loading conditions. DFT-MD is seen to provide provide accurate descriptions of Hugoniot state properties of dissimilar pure species as well as homogeneous mixtures of them.

\begin{acknowledgments}
We thank Dr. John Benage for valuable discussions on this work.  We thank Dr.~Joel Kress at Los Alamos National Laboratory for valuable discussions on mix rules and DFT/QMD simulations.  We thank Jesse Lynch and Nicole Cofer for assembling the cryo-targets and we thank Andrew Lopez, Keegan Shelton, and Jose Villalva for operating the cryogenics systems on Z.  The NNSA Science Campaigns supported this work. Sandia National Laboratories is a multi-program laboratory managed and operated by Sandia Corporation, a wholly owned subsidiary of Lockheed Martin Corporation, for the U.S. Department of EnergyÕs National Nuclear Security Administration under contract DE-AC04-94AL85000.
\end{acknowledgments}

\appendix
\section{Reshock States}
Although we did not calculate the reshock states using DFT, the target design shown in Fig.~\ref{ExptSchematic}, with the rear quartz top-hat, does permit the measurement of the reshock state in the ethane and mixture samples.  Using the method described in Ref.~\citenum{RootCO2:2013} the reshock state can be determined.  The method utilizes a fit to the $U_S$-$U_P$ data to determine the initial state of the sample prior to reshock because some attenuation can exist in the shock velocity as the shock traverses the sample.  Since we only have a few experimental points for the principal Hugoniot of ethane and the mixtures, we include the DFT data in the $U_S - U_P$ fits.  We assume a 0.5\% uncorrelated uncertainty in the DFT $U_S$-$U_P$ data points. The final state pressure $P$ and $U_P$ are known to a high degree of accuracy because of the quartz Hugoniot.\cite{MDKQuartz09}  The linear $U_S$-$U_P$ parameters along with the correlation between the fit parameters are listed in Table~\ref{Tab:FitParams}. The reshock states are listed in Tables~\ref{Tab:EthReshock} and \ref{Tab:MixtureReshock}.

\bibliographystyle{apsrev4-1}
\bibliography{XeEth_Mix}

\begin{table*}
  \addtolength{\tabcolsep}{6pt}
\renewcommand{\arraystretch}{0.7}
\caption{Linear fit parameters to the $U_S$-$U_P$ experimental data and DFT results. $U_S = C_0 + S_{1}U_P$.  The term $\sigma_{C_0}\sigma_{S_1}$ is the off-diagonal term in the covariance matrix of the fit parameters}
\begin{tabular}{ccccc}
\hline
   	   		& 	Range (km/s) 		&    $C_0$ (km/s)  		&   $S_1$   			&  $\sigma_{C_0}\sigma_{S_1}\times10^3$  		  \\  \hline \hline
Ethane 		& $U_S >$ 15.0		&  1.045$\pm$0.254		&  1.349 $\pm$ 0.017   	&     -4.1634  \\
Mass Mix 		& $U_S >$ 13.0		&  1.069$\pm$0.195		&  1.336  $\pm$ 0.014   	&    -2.6083 \\
Molar Mix 		& $U_S >$ 10.0		&  0.929$\pm$0.127	         &  1.313  $\pm$ 0.014   	&    -1.7045 \\
\hline
\end{tabular}
\label{Tab:FitParams}
\end{table*}

\begin{table*}
  \addtolength{\tabcolsep}{4pt}
\renewcommand{\arraystretch}{0.7}
\caption{Reshock data for liquid ethane.}
\begin{tabular}{ccccc}
\hline
   Shot & 	Ethane $U_S$  	&    Quartz $U_S$   		&   $\rho_2$   			&  $P_2$  		  \\ 
            &         (km/s)       		&       (km/s)      		&    (g/cm$^3$)    		& (GPa)  \\ \hline\hline
Z2277 & 24.26$\pm$0.06		&  18.36$\pm$0.04		&  2.864 $\pm$ 0.064   &     514.8 $\pm$  2.7 \\
Z2226 & 21.14$\pm$0.06		&  16.41$\pm$0.06		&  2.722  $\pm$ 0.068   &    397.9 $\pm$  3.4  \\
Z2331 & 19.81$\pm$0.05 		&  15.54$\pm$0.05	         &  2.684  $\pm$ 0.064   &    351.3 $\pm$  2.6 \\
\hline
\end{tabular}
\label{Tab:EthReshock}
\end{table*}

\begin{table*}
  \addtolength{\tabcolsep}{4pt}
\renewcommand{\arraystretch}{0.7}
\caption{Reshock data for the ethane-xenon mixtures.}
\begin{tabular}{cccccc}
\hline
   Shot & 	Mix Ratio 			&Mix $U_S$  	&    Quartz $U_S$   		&   $\rho_2$   			&  $P_2$  		  \\ 
           &	$x$ 				&         (km/s)       		&       (km/s)      		&    (g/cm$^3$)    		& (GPa)  \\ \hline\hline
Z2226 & 0.19				&   16.86$\pm$0.07		&  17.72$\pm$ 0.04  		&     6.347$\pm$0.430	& 474.5$\pm$2.5 \\
Z2277 & 0.19				&   19.12$\pm$0.08		&  19.88$\pm$ 0.05  		&     6.332$\pm$0.400	& 617.9$\pm$3.7 \\
Z2295 & 0.19				&   19.51$\pm$0.07		&  20.19$\pm$ 0.03  		&     6.380$\pm$0.406	& 640.3$\pm$2.3 \\
\hline
Z2527 (N)& 0.5				&   25.19$\pm$0.07		&  21.90$\pm$ 0.09  		&     4.379$\pm$0.185	& 771.8$\pm$7.4 \\
Z2527 (S)& 0.5				&   26.52$\pm$0.09		&  22.88$\pm$ 0.08  		&     4.405$\pm$0.185	& 853.5$\pm$7.0 \\

\hline
\end{tabular}
\label{Tab:MixtureReshock}
\end{table*}

 \end{document}